\documentclass[aps,prd,twocolumn,preprintnumbers,floatfix,nofootinbib,superscriptaddress]{revtex4-1}

\usepackage[utf8]{inputenc}
\usepackage[dvips]{graphicx}
\usepackage[dvipsnames]{xcolor}
\usepackage{color}
\usepackage{relsize}
\usepackage{graphics}
\usepackage{epstopdf}
\usepackage{hyperref}
\usepackage{mathrsfs}
\usepackage{amssymb}
\usepackage{physics}
\usepackage{booktabs}

\usepackage[normalem]{ ulem }
\usepackage{amsthm}
\usepackage{amsmath}
\usepackage{cancel}
\usepackage{fontawesome} 
\usepackage[caption=false]{subfig}
\usepackage{tabularx,ragged2e} 
 \newcolumntype{L}{>{\RaggedRight\arraybackslash}X}

\usepackage{comment}
\usepackage{array}
\usepackage[compat=1.1.0]{tikz-feynman}

\usepackage{hyperref}

\newcommand{\be}{\begin{equation}}
\newcommand{\ee}{\end{equation}}
\newcommand{\bea}{\begin{eqnarray}}
\newcommand{\eea}{\end{eqnarray}}
\newcommand{\ba}{\begin{aligned}}
\newcommand{\ea}{\end{aligned}}

\newcommand{\gs}{g_\star}
\newcommand{\gss}{g_{\star s}}

\newcommand{\mpsi}{m_{\psi}}
\newcommand{\mphi}{m_{\varphi}}
\newcommand{\mchi}{m_{\chi}}

\newcommand{\Imm}{\operatorname{Im}}

\newcommand{\magn}[1]{\lvert #1 \rvert}

\newcommand{\equaref}[1]{Eq.~(\ref{#1})}
\newcommand{\equasref}[2]{Eqs.~(\ref{#1})~and~(\ref{#2})}

\newcommand{\figref}[1]{Fig.~\ref{#1}}

\newcommand{\refref}[1]{Ref.~\cite{#1}}

\newcommand\myshade{80}
\colorlet{mylinkcolor}{ForestGreen}
\colorlet{mycitecolor}{Red}
\colorlet{myurlcolor}{violet}

\hypersetup{
 linkcolor = mylinkcolor!\myshade!black,
 citecolor = mycitecolor!\myshade!black,
 urlcolor = myurlcolor!\myshade!black,
 colorlinks = true
}

\usepackage{pgfplots}

\begin{document}

\preprint{IPMU23-0027}

\vspace*{1mm}

\title{Structure Formation after Reheating: Supermassive Primordial Black Holes and Fermi Ball Dark Matter}

\author{Marcos M. Flores}
\affiliation{Department of Physics and Astronomy, University of California, Los Angeles\\
Los Angeles, California, 90095-1547, USA}

\author{Yifan Lu}
\affiliation{Department of Physics and Astronomy, University of California, Los Angeles\\
Los Angeles, California, 90095-1547, USA}

\author{Alexander Kusenko}
\affiliation{Department of Physics and Astronomy, University of California, Los Angeles\\
Los Angeles, California, 90095-1547, USA}
\affiliation{Kavli Institute for the Physics and Mathematics of the Universe (WPI), UTIAS\\
The University of Tokyo, Kashiwa, Chiba 277-8583, Japan}
\date{\today}

\begin{abstract}
In the presence of (relatively) long-range forces, structures can form even during the radiation dominated era, leading to compact objects, such as Fermi balls or primordial black holes (PBHs), which can account for all or part of dark matter.  We present a detailed analysis of a model in which fermions are produced from the inflaton decay developing some particle-antiparticle asymmetry. 
These fermions undergo clustering and structure formation driven by a Yukawa interaction. The same interaction provides a cooling channel for the dark halos via scalar radiation, leading to rapid collapse and the formation of a compact object. We discuss the criteria for the formation of either PBHs and Fermi balls. In the PBH formation regime, supermassive PBHs can seed the active galactic nuclei or quasars found at high redshift. Alternatively, Fermi balls can account for all of the cold dark matter, while evading microlensing constraints.
\end{abstract}

\maketitle

\section{Introduction}


The structures in some species can grow even during the radiation dominated era in the presence of relatively long-range forces~\cite{Gradwohl:1992ue, Amendola:2017xhl, Savastano:2019zpr, Domenech:2021uyx, Flores:2020drq, Flores:2021jas, Flores:2021tmc, Flores:2022oef, Flores:2022uzt, Domenech:2023afs, Flores:2023nto}. One example is Yukawa interactions  which can form halos of interacting fermions~\cite{Flores:2020drq,Domenech:2023afs}. The fate of such halos depends on their cooling.  In the absense of dissipation, these halos would remain intact until the present day, or until their constituent particles decay. However, the same Yukawa interaction also introduces a dissipation channel in the form of scalar radiation. These scalar waves allow for the rapid removal of energy and angular momentum from the halo~\cite{Flores:2020drq, Flores:2021tmc}. Consequently, the halo can collapse to highly condensed fermionic matter, leading to either a compact Fermi ball or a primordial black hole (PBH). 

PBHs are fascinating objects that could form in the early Universe. They were initially conjectured to be formed from the collapse of large overdensities \cite{Hawking:1971ei, Carr:1974nx, Carr:1975qj} in the primordial power spectrum. Over the years, other formation mechanisms have been proposed. For example, PBHs could arise from the unstable scalar field condensate formed after inflation, such as Q balls \cite{Cotner:2016cvr, Cotner:2017tir, Cotner:2019ykd} or oscillons \cite{Cotner:2018vug}. Phase transitions in the dark sector can lead to the collapse of vacuum bubbles and produce PBHs even after photon decoupling \cite{Lu:2022jnp}. Unlike the formation scenario from critical collapse \cite{Khlopov:2008qy, Choudhury:2023jlt, Choudhury:2023rks}, the PBH (and Fermi ball) mass function in our model does not depend on the details of the underlying inflation theory.

Early structure formation does allow for the formation of halos composed of both fermions and antifermions. In fact, there are some cases in which the lack of an asymmetry might be useful~\cite{Flores:2022oef, Flores:2023nto}. Generically however, the formation of compact objects requires a particle-antiparticle asymmetry to avoid pair annihilation. Generation of a dark matter asymmetry has been explored in many contexts, with a particular focus on addressing the coincidence problem, i.e., the fact that the dark matter abundance is remarkably close to the abundance of ordinary  matter~\cite{Petraki:2013wwa, Zurek:2013wia}.

In this work, we will focus on the early structure formation with a dark sector asymmetry generated  through inflaton decay. As we will describe, depending on the temperature at the onset of early structure formation, compact objects can be formed in two classes of time; either instantaneously or noninstantaneously.  The resulting compact object can span a wide range of masses, including sub-Planck mass Fermi balls up to supermassive black holes (SMBH).  Finally, we will discuss possible consequences in cosmology -- the produced supermassive PBHs could seed the high redshift active galactic nuclei (AGN) observed by JWST \cite{Bunker:2023lzn}, while all of cold dark matter (CDM) can exist in the form of Fermi balls. 

\section{Asymmetry in the Dark Sector}

Let us consider a dark sector with three generations of fermions $\psi_i$, $\bar{\psi}_i$ ($i=1, 2, 3$) and a light scalar $\chi$. We work in the basis where the fermion mass matrix is diagonalized and, for simplicity, we assume the fermions have a common mass $\mpsi$. In addition, we consider three generations of Majorana fermions $N_i$ with a common mass $M$. If we take these Majorana fermions to be right-handed neutrinos, the decay of $N_i$'s can also generate a net lepton number \cite{Fukugita:1986hr}. Inflation is driven by a complex scalar field $\varphi$. We impose a global $U(1)$ symmetry on $\varphi$ and the dark sector. The charge assignment is as following:
\begin{center}
    \begin{tabularx}{0.3\textwidth} { 
  | >{\centering\arraybackslash}X 
  | >{\centering\arraybackslash}X 
  | >{\centering\arraybackslash}X
  | >{\centering\arraybackslash}X |}
 \hline
  & $\varphi$ & $\psi$ & $\bar{\psi}$ \\
 \hline
 $U(1)_D$  & -1 & 1 & -1 \\
\hline
\end{tabularx}
\end{center}

The Lagrangian of the dark sector and inflaton is given by
\begin{equation}
    \begin{aligned}
        \mathcal{L}_{\rm ds} =& \frac{1}{2} (\partial_{\mu}\chi)^2-\frac{1}{2} \mchi^2 \chi^2 - \mpsi \bar{\psi}_i\psi_i - y_{ij} \chi \bar{\psi}_i\psi_j  + h.c. \\
        \mathcal{L}_{\rm inf} =& \partial_{\mu}\varphi^{\dagger} \partial^{\mu}\varphi - \mphi^2 \magn{\varphi}^2  - V(\varphi) - \frac{1}{2} M N^c_i N^c_i \\
        & + \kappa_{ij} \varphi N^c_i \psi_j + \bar{\kappa}_{ij} \varphi^{\dagger} N^c_i \bar{\psi}_j  + h.c.
    \end{aligned}
\label{eq:Lagrangian}
\end{equation}
Here, $V(\varphi)$ is the potential that drives inflation and all the fermions in this model are two component spinors. We note that structure formation under the Yukawa potential in the dark sector is independent of the predictions from the inflation theory. Specifically, our model do not rely on a particular form of the inflation potential or the primodial power spectrum. 

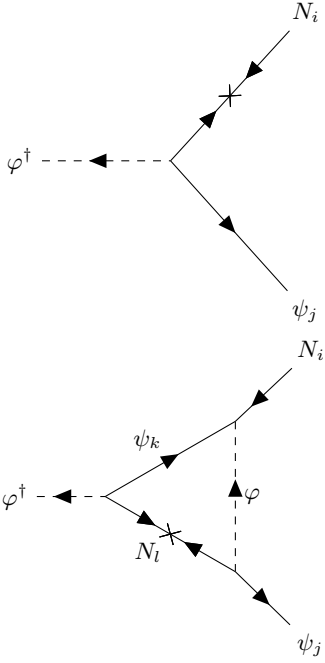
\begin{figure}
    \centering
        \begin{tikzpicture}[baseline=(a.base)]
            \begin{feynman}
                \vertex (i) {\(\varphi^{\dagger}\)};
                \vertex [right=2cm of i] (a);
                \vertex [above right=1.73cm and 1.5cm of a] (f1) {\(N_i\)};
                \vertex [below right=1.73cm and 1.5cm of a] (f2) {\(\psi_j\)};

                \diagram* {
                    (i)  -- [anti charged scalar] (a),
                    (a) -- [majorana, insertion=0.5] (f1) ,
                    (a) -- [fermion] (f2) ,
                    };
            \end{feynman}
        \end{tikzpicture}
        \quad \quad \quad
        \begin{tikzpicture}[baseline=(a.base)]
            \begin{feynman}
                \vertex (i) {\(\varphi^{\dagger}\)};
                \vertex [right=1.2cm of i] (a);
                \vertex [above right=1cm and 1.73cm of a] (b);
                \vertex [below right=1cm and 1.73cm of a] (c);
                \vertex [above right=1cm of b] (f1) {\(N_i\)};
                \vertex [below right=1cm of c] (f2) {\(\psi_j\)};

                \diagram* {
                    (i)  -- [anti charged scalar] (a),
                    (a) -- [fermion, edge label=\(\psi_k\)] (b),
                    (b) -- [anti charged scalar, edge label=\(\varphi\)] (c),
                    (c) -- [majorana, insertion=0.5, edge label=\(N_l\)] (a),
                    (b) -- [anti fermion] (f1) ,
                    (c) -- [fermion] (f2) ,
                    };
            \end{feynman}
        \end{tikzpicture}
    \caption{Diagrams of inflaton decay at tree and one loop levels.}
    \label{simplediagram}
\end{figure}

Reheating is driven by the inflaton decay to the dark sector via $\varphi \rightarrow N_i + \bar{\psi}_j$ and the subsequent decay of $N_i$'s which also generates the Standard Model (SM) plasma. Since the mass matrices are all diagonalized, the couplings $\kappa_{ij}$ and $\bar{\kappa}_{ij}$ are in general complex. This leads to a different decay rate for the charge conjugated decay mode. Due to the $\psi$ number violating interactions, an asymmetry in the dark sector will be generated after reheating. The asymmetry generated per decay can be calculated from the interference of the diagrams in \figref{simplediagram} (and the corresponding diagrams for the decay of $\varphi$). We note that this calculation is reminiscent of similar scenarios encountered in leptogenesis~\cite{Fukugita:1986hr, Davidson:2008bu}.
We find the result to be
\begin{equation}
    \varepsilon = \frac{2\ln 2 -1}{8\pi}\left(\frac{M}{\mphi}\right)^2 \frac{\Imm (\xi - \bar{\xi})}{\magn{\kappa_{ij}}^2 + \magn{\bar{\kappa}_{ij}}^2}, 
    \label{asymsim}
\end{equation}
where
\begin{equation}
    \xi = \kappa_{ij}^{\ast}\kappa_{ik}^{\ast}\kappa_{lj}\kappa_{lk},
\end{equation}
and $\bar{\xi}$ is obtained by replacing $\kappa$ with $\bar{\kappa}$. In general, the couplings $\kappa_{ij}$ and $\bar{\kappa}_{ij}$ are independent of each other, so the asymmetry produced by the decay of $\varphi$ will not be canceled by the decay of $\varphi^{\dagger}$. Observe that in~\equaref{asymsim}, the fermion mass suppression factor $(M / \mphi)^2$ will lead to final asymmetry which is expected to be very small.

The evolution of the $\psi$ number density $n = n_{\psi} - n_{\bar{\psi}}$ during reheating follows the Boltzmann equation
\begin{equation}
    \dot{n} + 3 H n = \varepsilon \Gamma \frac{\rho_{\varphi}}{\mphi} - \frac{3 K T^5}{4 \pi^3 \mphi^4} n,
    \label{simbol}
\end{equation}
with
\begin{equation}
    K = \magn{\kappa_{ij}\bar{\kappa}_{kl}}^2 + \magn{\bar{\kappa}_{il}\kappa_{jk}}^2.
\end{equation}

In \equaref{simbol}, the first term on the right-hand side corresponds to the production of asymmetry from the inflaton decay. Washout of the generated asymmetry will also occur due to the $\psi \psi \rightarrow N N$ and $\psi N \rightarrow \bar{\psi} N$ interactions which are mediated by the inflaton. This leads to an overall asymmetry which will be suppressed. The effect of washout is encapsulated in the second term on the right-hand side of \equaref{simbol}. The temperature $T$ is the plasma temperature during reheating, which we take to be the same between the dark sector and SM sector. The radiation density and the inflaton energy density follow the equations:
\begin{equation}
    \begin{aligned}
        \dot{\rho}_{\varphi} + 3 H \rho_{\varphi} &= - \Gamma \rho_{\varphi}, \\
        \dot{\rho}_R + 4 H \rho_R &= \Gamma \rho_{\varphi},
    \end{aligned}
    \label{energydensity}
\end{equation}
where $\rho_R = \gs \pi^2 T^4/30$ and $\gs = 123.5$ is the degrees of freedom of the SM and the dark sector including three generations of Majorana fermions. The inflaton decay rate is given by
\begin{equation}
    \Gamma = \frac{1}{8 \pi} (\magn{\kappa_{ij}}^2 + \magn{\bar{\kappa}_{ij}}^2) \mphi.
\end{equation}

\begin{figure}
    \centering
    \includegraphics[width=0.48\textwidth]{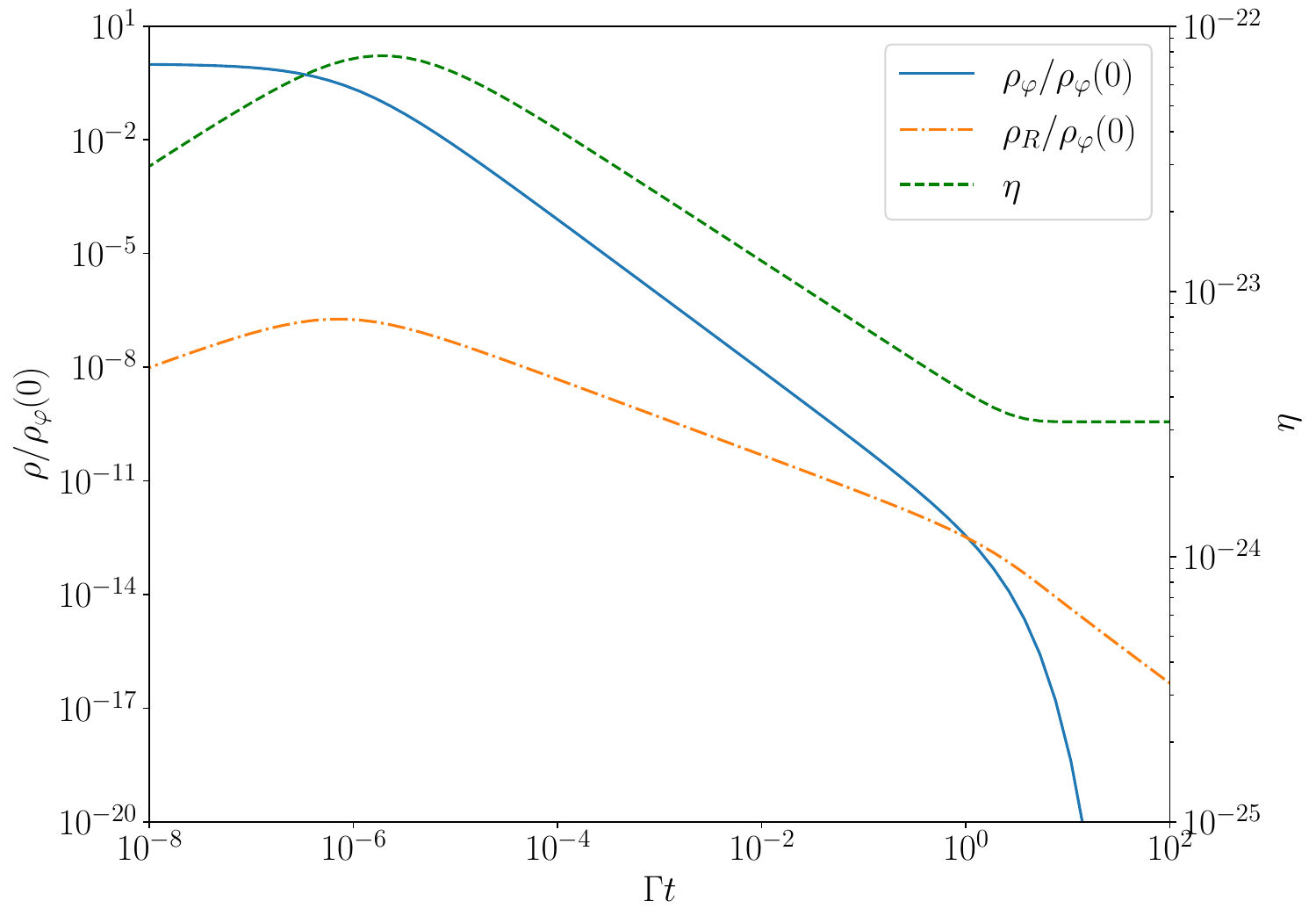}
    \caption{Inflaton, radiation energy density and asymmetry as functions of time. The parameters are chosen to be $\kappa = 2\times10^{-4}$, $\mphi = 1\times10^{16}$ GeV, $\varepsilon = 1\times10^{-20}$ and the initial inflaton energy is taken to be $1\times 10^{16}$ GeV.}
    \label{asymmetry}
\end{figure}

The $\psi$ asymmetry produced is commonly parametrized by $\eta = n/s$, where $s$ is the entropy density of the ambient plasma. At the beginning of reheating, both of asymmetry production and washout interactions are active due to the high plasma temperature. As the temperature decreases during reheating, the washout interactions gradually freeze out and a constant asymmetry $\eta$ is produced near the end of inflaton decay. If we neglect the washout terms, then the final asymmetry produced will be $\eta \sim \epsilon$. In order to compute the final asymmetry including the washout effects, we numerically solve \equasref{simbol}{energydensity} by taking $\kappa_{ij} \sim \bar{\kappa}_{ij} \equiv \kappa$ for all $i, j$. Due to the mass suppression $M/\mphi$, we can choose the value of the parameter $\epsilon$ independently from $\kappa$ and require $\epsilon \ll \magn{\kappa}^2$. In addition, the coupling $\kappa$ should also be sufficiently suppressed such that the flatness of the inflaton potential is preserved at the quantum level. An example of the time evolution of the energy densities and the asymmetry is shown in Fig.~\ref{asymmetry}. We see that indeed a constant asymmetry $\eta$ is produced. As expected, washout effects reduce the value $\eta$ to be much smaller than the parameter $\epsilon$.

Since $\varphi$ carries a dark quantum number, fluctuations of the
complex inflaton could generate isocurvature perturbations in the dark sector. As we will show later on, the dark sector fermions will collapse into Fermi balls which provide all of the CDM that we observe today. This translates to CDM isocurvature fluctuations that are constrained by Planck \cite{Planck:2018jri}. However, this process depends on the inflation trajectory and possible additional $U(1)_D$ violating operators in the inflaton potential. So the analysis of this effect is tangential to the main purpose of this paper. A detailed calculation is done in the case of baryogenesis in \refref{Cline:2019fxx} and a two-field inflation model is allowed for a suitable choice of the inflaton potential $V(\varphi)$.

\section{Structure formation mechanism}

The inclusion of light scalars $\chi$ in~\equaref{eq:Lagrangian} introduces an instability in the $\psi$ fluid, analogous to the gravitational instability in conventional structure formation. Gravity alone does not permit for the growth of structure during the earliest epochs of the Universe. However, the inclusion of Yukawa forces, which are generally stronger than gravity, can generate structure on timescales significantly faster than the Hubble time -- even during radiation domination. This principle of early structure formation has been examined in many contexts~\cite{Gradwohl:1992ue, Amendola:2017xhl, Savastano:2019zpr, Domenech:2021uyx, Flores:2020drq, Flores:2021jas, Flores:2021tmc, Flores:2022oef, Flores:2022uzt, Domenech:2023afs, Flores:2023nto}.

In its simplest form, early structure formation involves a heavy fermion interacting via a Yukawa interaction. The growth of $\psi$ overdensities $\Delta_k\equiv (n_\psi - \bar{n}_\psi)/\bar{n}_\psi$ is described by the differential equation~\cite{Domenech:2023afs},
\begin{equation}
\label{eq:OverdensityDE}
\Delta_k''
+
\frac{2 + 3x}{2x(1 + x)}\Delta_k'
=
\frac{3}{2x(1 + x)}\Delta_k f_\psi
\left[
1 + 
\frac{2\beta^2}{1 + (k\ell)^{-2}}
\right]
\end{equation}
where $\prime\equiv d/dx$ with $x = a/a_{\rm eq}$ being the scale factor normalized to the time where $\rho_R = \rho_m + \rho_\psi$ and lastly, $f_\psi = \rho_\psi/(\rho_m + \rho_\psi)$. Here, $\beta$ characterizes the strength of the force relative to gravity and $\ell$ is the distance over which the Yukawa interaction is relevant. For simplicity, in our analysis we assume $\beta$ is a constant given by $\beta = yM_{\rm Pl}/m_\psi$ and $\ell = m_\chi^{-1}$, though in general, additional terms in the $\chi$ potential can induce a density-dependent $\beta$ and $\ell$~\cite{Domenech:2023afs}.

In the limit of a massless mediator, i.e. $\ell \to \infty$, the $k$ dependence from Eq.~\eqref{eq:OverdensityDE} is eliminated. In massless mediator limit, the growing mode from Eq.~\eqref{eq:OverdensityDE} is given by
\begin{equation}
    \Delta_k(t) = \Delta_k (0) I_0 \left(4 \sqrt{p}\left(t / t_{\mathrm{eq}}\right)^{1 / 4}\right),
\end{equation}
where
\begin{equation}
    p=\frac{3}{8}\left(1+\beta^2\right),
\end{equation}
and $I_0$ is the modified Bessel function of the first kind. This solution exhibits exponential growth behavior as long as $\sqrt{p}(t / t_{\rm eq})^{1/4} \gg 1$. This constrains the temperature for rapid structure formation in the radiation dominated era to be 
\begin{equation}
    T \ll \left(\frac{g_{\star s, \rm eq}}{\gss}\right)^{1/3}\beta^2 T_{\rm eq} \equiv T_{\rm form}.
\end{equation}

For simplicity, we only require $T < T_{\rm form}$ in the following analysis. In addition to this restriction, we also require that $\bar{\psi}\psi\leftrightarrow \chi\chi$ interactions to have frozen out, so that halos may be generated from a stable population of particles. Lastly, we will also require that early structure formation occurs once $T\lesssim m_\psi/3$. These three conditions allow us to define the ``growth temperature"~\cite{Flores:2020drq},
\begin{equation}
    T_g = 
    \min\left(
    \frac{\mpsi}{3}, \frac{\mpsi}{ \ln{(y^4 m_p/\mpsi)}}, T_{\rm form}
    \right),
    \label{Tg}
\end{equation}
at which early structure formation begins. For the remainder of this work, we focus on structure formation after reheating so our model is valid only for $T_g < T_{\rm RH}$. The possibility of structure formation during the inflaton decay is also interesting and will be left for future work.

The subsequent evolution of the halos depends on the range of the Yukawa interaction $1 / \mchi$ and the horizon size $H_{g}^{-1}$ at $T_g$. If $\mchi \gg H_g$, then structure formation occurs rapidly at $T_g$ and the subhorizon sized halos quickly undergo scalar radiation cooling. As we will discuss shortly, these subhorizon halos can radiatively collapse, and depending on the parameters under consideration, result in either PBHs or Fermi balls. Since the timescale for halo formation is much smaller than the Hubble time, we refer to this scenario as ``instantaneous formation.'' If $\mchi \ll H_g$, then subhorizon sized halos form initially. However, as the horizon expands, Yukawa force can also drive halo mergers, which we expect to be very efficient compared to gravitational mergers. The maximum mass of the halo is determined by the time when the length scale $1/\mchi$ enters the horizon. This happens at a temperature lower than $T_g$ and we refer to this process as ``noninstantaneous formation."

The overdensities described by \equaref{eq:OverdensityDE} quickly grow to exceed unity, thus entering the nonlinear regime. At this stage, the overdenisites evolve into virialized $\psi$ halos. In the absence of dissipation, these virialized halos would simply remain as such, unless the constituent particles themselves can decay. However, the same Yukawa interaction which generated overdensities in the $\psi$ fluid also enables the emission of scalar radiation from accelerating $\psi$ particles. Initially, when the halo is sufficiently dilute, energy is primarily dissipated through pair interactions, i.e., through free-free or bremmstrahlung emission. As a given $\psi$ halo contracts, its density will increase until $\chi$ radiation becomes trapped. In this regime, radiation is only emitted from the surface of the halo. For radiative collapse to lead to the formation of a compact object, the characteristic timescale associated with cooling,
\begin{equation}
\tau_{\rm cool}
\equiv
\frac{E}{dE/dt}
=
\frac{E}{P_{\rm brem} + P_{\rm surf} + \cdots}
\end{equation}
must be shorter that the Hubble time $H^{-1}$. Since all halos must traverse the inefficient surface cooling regime before reaching their final form, the surface cooling timescale dictates whether a halo can collapse quickly enough. 
Using the Stefan-Boltzmann law, we find that $\tau_{\rm cool} \sim \tau_{\rm surf} \sim R_h$. For $m_\chi\gg H$, the maximum halo radius $R_h \sim m_\chi^{-1} \ll H^{-1}$ implying that subhorizon halos will cool rapidly compared to the Hubble time. This process corresponds to the ``instantaneous formation" regime discussed above.

Alternatively, when the mediator mass is smaller than the Hubble parameter, subhorizon halos will form, but these smaller halos will quickly merge until the combined halos produce a larger halo of radius $R_h\sim H^{-1}$. This process continues until $m_\chi \sim H$. Once halos become small relative to the horizon size, surface cooling will be efficient enough to drive them toward collapse. This process corresponds to the ``noninstantaneous formation" regime discussed previously. 

These different scenarios will each lead to different constraints on parameter space. Therefore, we will tackle the each situation individually.

\subsection{Instantaneous formation}

The instantaneous formation scenario relies on the mediator mass, $m_\chi$, being larger than the Hubble parameter as structures begin to grow. This sets a strict lower bound on the mediator mass,
\begin{equation}
    \mchi > \sqrt{\frac{4 \pi^3 \gs}{45}} \frac{T_g^2}{m_p}.
    \label{instcondition}
\end{equation}
Under this restriction, the halo mass is well approximated by the total $\psi$ mass within the $1/\mchi$ radius: 
\begin{equation}
    M_h = \frac{4\pi}{3} R^3 \rho_{\psi} = \frac{8 \pi^3 \gss}{135} \frac{\eta \mpsi T_g^3}{\mchi^3}.
    \label{PBHmass}
\end{equation}

At the final stage of the halo collapse, the halo is reheated. Subsequently, the washout effect of the dark sector asymmetry, caused by the $\psi$ number violating scattering, is reactivated. This reactivation has the potential to annihilate the existing asymmetry within the collapsing halo. The timescale of this process is determined by the washout terms in \equaref{simbol}: $\tau_{\mathrm{ann}}^{-1} \sim \kappa^4 T^5_{h} / m^4_{\varphi}$. For successful PBH or Fermi ball formation, the $\psi$ particles inside the halo cannot be totally annihilated before the end of collapse. This requires the collapse timescale $\tau_c < \tau_{\mathrm{ann}}$. We estimate $\tau_c$ as the following: For simplicity, we assume that $y_{ij} = y$ among all generations. By the virial theorem,
\begin{equation}
    \frac{4\pi^4}{90} R^3 T_h^4 = \frac{y^2 N^2}{R},
\end{equation}
which implies the temperature of the halo increases as $1/R$. In the Yukawa energy term, we neglected possible order 1 factors that comes from the radial density profile or non-spherical distribution of matter in the halo. Annihilation is only efficient at high halo temperatures, and radiation cooling at this stage is dominated by the surface radiation channel:
\begin{equation}
    \frac{dE}{dt} = 4 \pi R^2 T_h^4 \sim \frac{y^2 N^2}{R^2} \frac{dR}{dt}.
\end{equation}
From this, we can calculate the collapse timescale $\tau_c = R / (dR/dt)$ by
\begin{equation}
    \tau_c = \frac{y^2 N^2}{R^3 T_h^4} \sim \frac{\sqrt{y N}}{T_h} \sim \frac{\sqrt{y \eta}}{T_h} \left(\frac{T_g}{\mchi}\right)^{3/2},
\end{equation}
where we again used the virial theorem. Comparing this with $\tau_{\rm ann}$, we obtain the following constraint:
\begin{equation}
    \left(\frac{\mchi}{T_g}\right)^{3/2} > \kappa^4 \sqrt{y \eta} \left(\frac{T_h}{\mphi}\right)^4,
    \label{annboundsim}
\end{equation}
which must be satisfied for all achievable halo temperature $T_h$. The most stringent case occurs at the radius $R_f$ of the final compact object. The maximum temperature is estimated to be 
\begin{equation}
    T_{\rm max} = \frac{T_g}{\mchi R_f}.
\end{equation}

Deducing the type of the compact object formed (PBH/Fermi ball) will require a more involved analysis and we will discuss this in detail in Sec. \ref{IIIC}. However, the qualitative behavior can be understood by comparing the scaling of $R_{f}$. For PBHs, the final radius is simply $R_f = 2 G M_h$ so $R_f \sim N$. Alternatively, the Fermi balls radius scales as $R_f \sim N^{2/3} / \mpsi$ (see \equaref{FBradius}). It follows then that the Schwarzschild radius of the halo will exceed the Fermi ball radius at large $N$. Therefore, our model generally predicts PBHs on the higher mass end while Fermi balls are formed on the lower mass end.

Finally, the Compton wavelength of $\psi$'s must be larger than $R_f$ for the halos to collapse. This is always satisfied for Fermi balls with $N \gg 1$ which can be seen from the scaling formula. We also explicitly check this condition in the PBH formation regime.

\subsection{Noninstantaneous formation}

In the noninstantaneous formation scenario, subhorizon sized Fermi balls are formed initially. As the Universe cools down, more Fermi balls enter the horizon and the Yukawa interaction continues to drive the structure formation through Fermi ball merges. There are two temperatures of importance in this scenario. The first is the growth temperature $T_g$, at which density perturbation grows and halos form, previously defined in \equaref{Tg}. The second relevant temperature, $T_c$, corresponds to the point in time at which mergers stop and halos start to radiatively cool and collapse. Based on our previous discussion, $T_c$ occurs at the temperature when the length scale $1/\mchi$ enters the horizon:
\begin{equation}
    T_c =  \left(\frac{45}{4 \pi^3 \gs}\right)^{1/4} \sqrt{\mchi m_p}.
    \label{noninstT}
\end{equation}

Much like conventional structure formation, we expect a rich history of mergers which steadily generate larger dark matter halos, albeit on a significantly shorter timescale. The exact mass function of such objects requires many-body simulation of the merger dynamics. Here, we simply approximate the halo mass function as monochromatic and peaked at $M_h$, where the halo mass $M_h$ is given by the total mass of all the available halos within the horizon at $T_c$. Since the halo mergers are driven by Yukawa interactions, which again, are significantly stronger than gravity, we expect mergers to be very efficient. The approximation is justified by recent $N$-body simulations which have illustrated the development of structure during radiation domination~\cite{Domenech:2023afs}. Crucially, these simulation show a marked absence of filimentary structure, indicating that nearly all of the heavy fermions end up in halos, with the halo's radius being determined by the mediator's Compton wavelength.

Here, we compute the halo mass and derive the constraints again as in the instantaneous formation scenario. The halo density before the mergers is given by
\begin{equation}
    n_h = \frac{1}{H_g^{-3}} \frac{s_c}{s_g},
\end{equation}
where $s_c$ and $s_g$ are the entropy densities at $T_c$ and $T_g$, respectively. Assuming that $\gss$ is almost a constant between $T_g$ and $T_c$, the total halo mass
\begin{equation}
    M_h = \eta \mpsi s_g H_g^{-3} \frac{4\pi n_h}{3 \mchi^3} = \frac{8 \pi^3 \gss}{135} \frac{\eta \mpsi T_c^3}{\mchi^3}.
    \label{noninsmass}
\end{equation}
The halo temperature at formation can also be estimated using the virial theorem:
\begin{equation}
    T_i = \left( \frac{90}{4 \pi^4}\right)^{1/4} \frac{\sqrt{y N}}{R} = \left( \frac{32 \pi^2 \gss^2}{405}\right)^{1/4} T_c \sqrt{\frac{y \eta T_c}{\mchi}}.
\end{equation}

The estimation of $\tau_c$ is slightly different in this scenario for the halo mass \equaref{noninsmass}, and we obtain
\begin{equation}
    \tau_c \sim \frac{\sqrt{y \eta}}{T_h} \left(\frac{T_c}{\mchi}\right)^{3/2}.
\end{equation}
So the annihilation constraint becomes
\begin{equation}
    \left(\frac{\mchi}{T_c}\right)^{3/2} > \kappa^4 \sqrt{y \eta} \left(\frac{T_h}{\mphi}\right)^4,
\end{equation}
with the maximum halo temperature given by
\begin{equation}
    T_{\rm max} = \left( \frac{32 \pi^2 \gss^2}{405}\right)^{1/4} \frac{\sqrt{y \eta}}{R_f} \left(\frac{T_c}{\mchi}\right)^{3/2}.
\end{equation}
The Compton wave length of $\psi$'s are always larger than the PBH/Fermi ball radius, as in the case of instantaneous formation.

Scalar cooling occurs alongside the Yukawa mergers and could lead to the formation of compact objects before mergers reach the $1/m_\chi$ threshold. To estimate this effect, we note that all mergers occur at the scale $\ll 1/ \mchi$. The potential due to a single halo is $V(r) \sim y^2 N/r$. The timescale for the initial mergers can be estimated by the infall time of the two initially static halos:
\begin{equation}
    t_{\rm infall} = \frac{\pi}{4} \frac{\mpsi d^{3/2} }{y M_{h}^{1/2}},
\end{equation}
where $d \sim H_g^{-1}$ is the average halo distance. For the cooling time, a conservative lower bound is given by the initial halo radius $\sim H_g^{-1}$. However, even if $t_{\rm infall} > \tau_{\rm cool}$, the initially formed compact objects will be Fermi balls of small mass which interact via the Yukawa interaction. Therefore, it is still possible to form objects with the final mass given by \eqref{noninsmass}.

\begin{figure}
         \centering
         \includegraphics[width=0.48\textwidth]{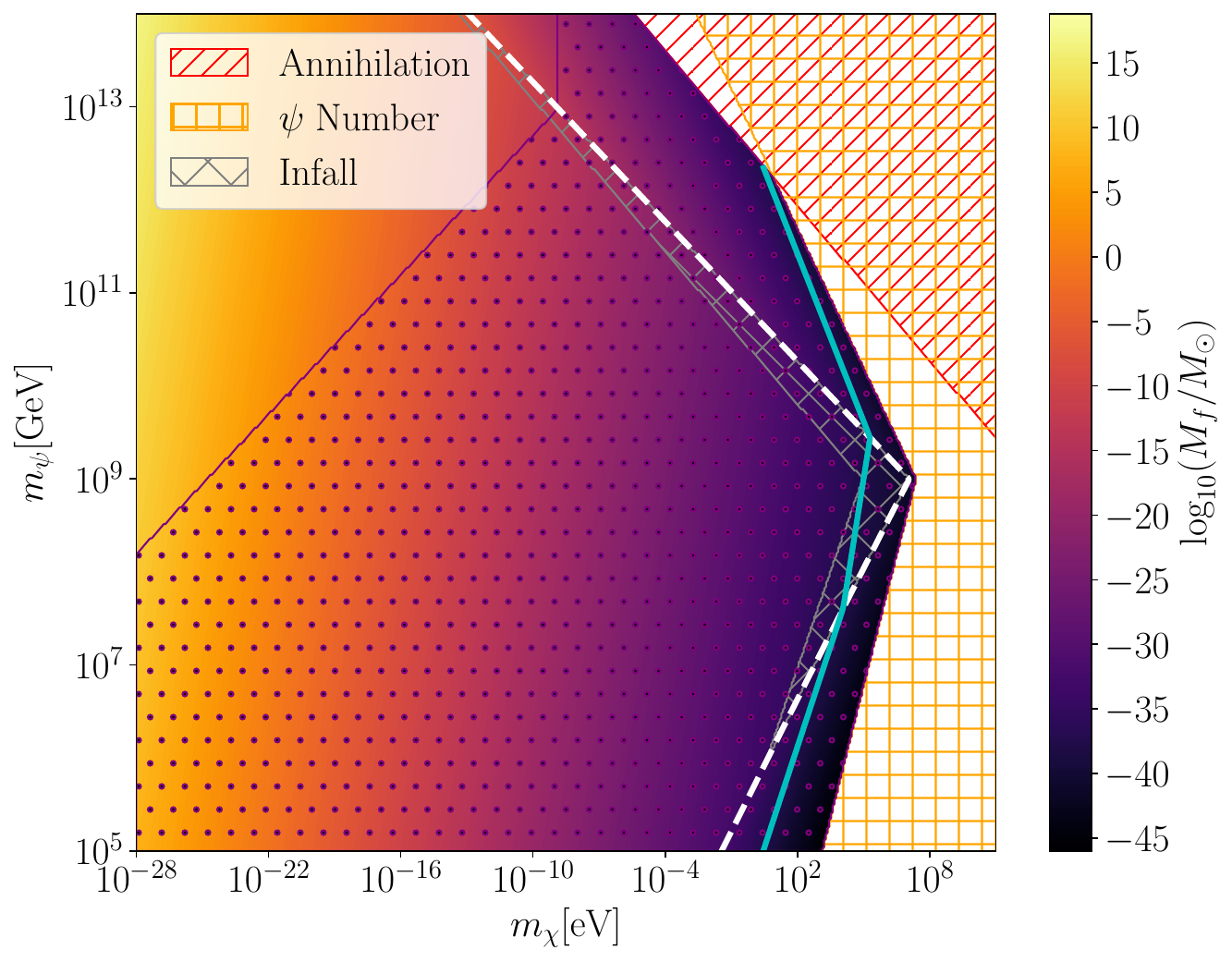}
         \caption{PBH/Fermi ball mass on the available parameter space. The purple grid marks the region where Fermi balls form and the instantaneous/noninstantaneous formation is separated by the white dashed line. The region covered by red and orange mesh is excluded by the annihilation and total particle number constraint (see Sec. \ref{IIIC}) and the gray mesh marks the region where cooling could be faster than mergers. The contour for Planck mass Fermi balls is shown by the solid cyan line. The parameters are chosen to be $\mphi = 10^{16}$ GeV, $y=5\times 10^{-2}$ and $\kappa = 2 \times 10^{-4}$ and $\eta =  10^{-25}$.
         } 
         \label{parameter}
\end{figure}

In \figref{parameter}, we demonstrate an example of the available parameter space and the PBH/Fermi ball mass distribution. The instantaneous formation scenario occurs for larger values of $\mchi$ (to the right of the white dashed line). For heavy $\psi$ particles ($\gtrsim 10^9$ GeV), the growth temperature $T_g$ is determined by the condition $T_g < T_{\rm form}$ whereas for lower $\psi$ mass, the growth of structure occurs immediately after the decoupling of $\psi$. This leads to the change in the boundary direction between instantaneous/noninstantaneous formation.

\subsection{The halo destiny}
\label{IIIC}

In the final stages of scalar cooling, compact objects with high fermion density are formed. The final state of these compact objects, whether it be as PBHs or Fermi balls, depends on the nuclear physics in the dark sector. At the final stage of the collapse, the dark fermions are reheated to become highly relativistic. The effective fermion mass 
\begin{equation}
    m_{\ast} = \mpsi + y \langle \chi \rangle
\end{equation}
are reduced by the nonzero vacuum expectation value (vev) of $\chi$ within the halo~\cite{Domenech:2021uyx}. This enhances the Fermi degeneracy pressure and an equilibrium state can be reached if the Yukawa attraction is balanced by this enhanced pressure. This is explored in Refs. \cite{Gresham:2017zqi, Gresham:2018rqo, Gresham:2017cvl} using relativistic mean field theory from the Walecka model \cite{Walecka:1974qa, Serot:1984ey} and also in Ref. \cite{DelGrosso:2023trq} as fermionic soliton solutions in general relativity. The qualitative behavior of the system is summarized as follows: At large $N$, when the radius of the Fermi ball is much larger than $1/ \mchi$, the system is relativistic and reaches the saturation limit where $R \sim N^{1/3}$. The stability of the resulting object is determined by solving the relativistic equation of state together with the Tolman-Oppenheimer-Volkof (TOV) equations in \refref{Gresham:2018rqo}. At medium $N$, the constituents remain relativistic but $R \ll 1/\mchi$. One has to take into account the nontrivial spatial variation of the scalar vev within the Fermi ball. The correct scaling of the radius is found to be $R \sim N^{2/3}$. This will be the relevant case in our scenario since the final compact objects are formed by radiative cooling of the initial $1/\mchi$ sized halos. In the limit $\mchi\rightarrow 0$, the stable radius is given by \cite{Gresham:2017zqi}
\begin{equation}
    R_{\rm FB} = \left(\frac{243}{2048\pi^2}\right)^{1/6} \frac{y N^{2/3}}{\mpsi}.
    \label{FBradius}
\end{equation}
For a finite $\mchi$, the saturation limit is reached for 
\begin{equation}
    N_{\rm sat} \sim \left( \frac{\mpsi}{y \mchi} \right)^{3/2}.
\end{equation}
In our model, 
\begin{equation}
    N = \frac{8 \pi^3 \gss}{135} \frac{\eta T^3}{\mchi^3},
\end{equation}
where $T = T_c$ or $T_g$. Therefore, \equaref{FBradius} is valid as long as
\begin{equation}
    T^3 \ll \frac{1}{\eta}\left( \frac{\mpsi \mchi}{y}\right)^{3/2}.
\end{equation}
Using Eqs. (\ref{instcondition}) or (\ref{noninstT}), one can show that it suffices to require $\eta \ll (\mpsi / y m_p)^{3/2}$ in both formation scenarios. This is easily satisfied as typically, $\eta < 10^{-20}$ and $\mpsi \gg 10^5$ GeV in our model. In addition, we require that $N > 10^{6}$ so that there are always enough particles for halo formation.  

In principle, the condition for PBH formation requires a TOV analysis of the stability of the system to take into account general relativistic effects. Here, we adopt a heuristic argument and declare that a PBH has formed when $R_{\rm FB} < 2 G M_f$. We leave the detailed analysis combining TOV equations as future work. The region of parameter space relevant for the formation of Fermi balls is marked by the purple grid in  \figref{parameter}.

\section{Significance in Cosmology}

In our scenario, either Fermi balls or PBHs are formed in the early radiation dominated Universe. These compact objects can be produced across a large span of masses. Of particular interest to this work will be {\it (i)} PBHs with masses $\sim 10^{5} - 10^{6}\ M_\odot$, i.e., SMBH and {\it (ii)} Fermi balls with masses $< 10^{-10} M_\odot$.

\subsection{Supermassive black holes}

\begin{figure}
         \centering
         \includegraphics[width=0.48\textwidth]{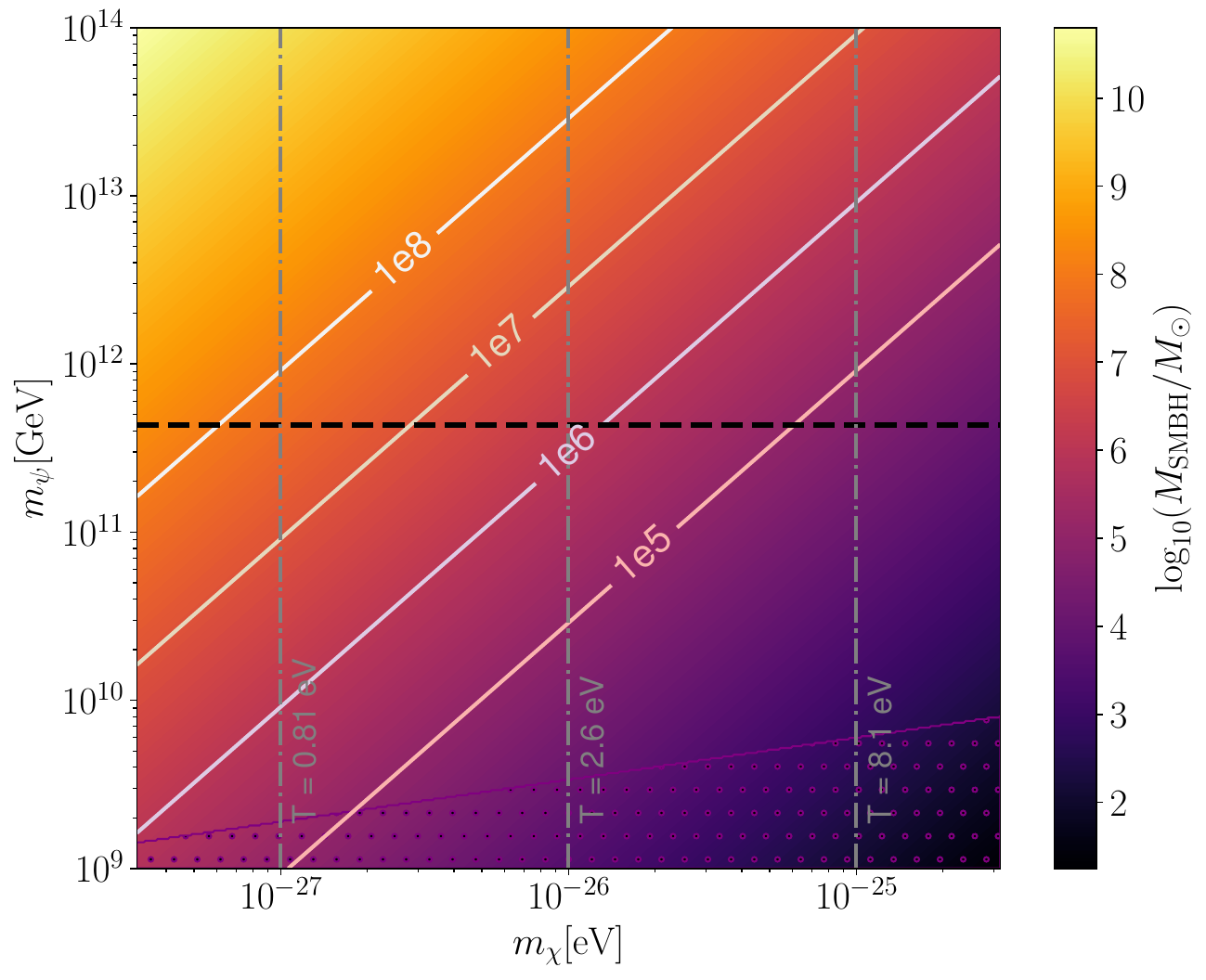}
         \caption{SMBH mass (in $M_\odot$) predicted from our model (solid contours). The black dashed line indicates the required abundance (with negligible accretion) by quasar observations. The purple grid marks the region where Fermi balls instead of PBHs form. The temperature at final halo formation (in the noninstantaneous scenario) is labeled on the dash-dotted lines. The parameters are chosen to be $\mphi = 10^{16}$ GeV, $y=5\times 10^{-2}$ and $\kappa = 2 \times 10^{-4}$ and $\eta = 10^{-30}$} 
         \label{massctr}
\end{figure}

PBHs have the ability to grow through accretion and therefore may seed high redshift AGNs \cite{Clesse:2015wea, Bean:2002kx}. This possibility has the potential to explain the apparent formation of galaxies at redshifts higher than predicted by $\Lambda\rm CDM$. This aspect is particularly interesting given the recent JWST observations of AGN at $z\sim 11$ \cite{Bunker:2023lzn}. In addition, PBHs could also act as progenitors of the SMBHs found in the quasars at $z \sim 6$ \cite{volonteri2011assessing, willott2010eddington}. However, we should note that unlike the ordinary formation scenario where PBHs are produced from the primordial density fluctuations, the abundance of PBHs can be produced in our model without requiring a highly non-Gaussian tail of the fluctuation \cite{Nakama:2017xvq, Nakama:2016kfq, Garcia-Bellido:2017aan} (see Refs. \cite{Garriga:2015fdk, Huang:2023chx} for other SMBH formation scenarios). Following the discussion in \refref{Serpico:2020ehh}, we note that the PBHs could either be SMBH at formation or intermediate mass black holes and undergo significant growth before redshift $z \sim 6$. In the later scenario, accretion near the Eddington limit is needed but a growth factor of $\sim 10^5$ is still achievable for the accretion model discussed in \refref{Serpico:2020ehh}. 

The observed AGNs at $z = 6$ determines the fraction of dark matter in PBHs above $10^6 M_\odot$ to be $f_{\rm PBH}\equiv \Omega_{\rm PBH}/\Omega_{\rm DM} \sim 2.9 \times 10^{-9}$ \cite{Serpico:2020ehh, volonteri2011assessing, willott2010eddington}. Due to theoretical uncertainties involving PBH accretion prior to $z=6$, we assume that the supermassive PBHs produced in our mechanism have negligible accretion rates before $z=6$. Even in this extreme case, the required PBH abundance is still consistent with the CMB bound \cite{Serpico:2020ehh, DeLuca:2020fpg}, which constrains the PBH with mass $\sim 10^5 M_\odot$ to have $f_{\rm PBH} \lesssim 10^{-9}$.

The abundance of PBHs in our scenario will rely on the fact that essentially all of the heavy fermions $\psi$ are captured into halos. This allows us to use the fermion energy density to determine the fraction of PBHs present today,
\begin{equation}
\begin{split}
f_{\rm PBH} &= \frac{\Omega_{\rm PBH}}{\rm DM}
    \approx 
    \frac{\Omega_{\psi}}{\Omega_{\rm DM}}\\[0.25cm]
    &\approx
    0.23 \left(\frac{\mpsi}{10^{10} \rm GeV}\right) \left(\frac{\eta}{10^{-20}}\right).
\end{split}
\end{equation}
The observed abundance can be produced by taking $\eta = 10^{-30}$. The corresponding scalar mass falls in the range of ultralight scalars. In \figref{massctr}, we labeled several contours that produce the SMBH mass of interest.

\subsection{Fermi balls as dark matter candidates}

\begin{figure}
    \centering
    \includegraphics[width=0.48\textwidth]{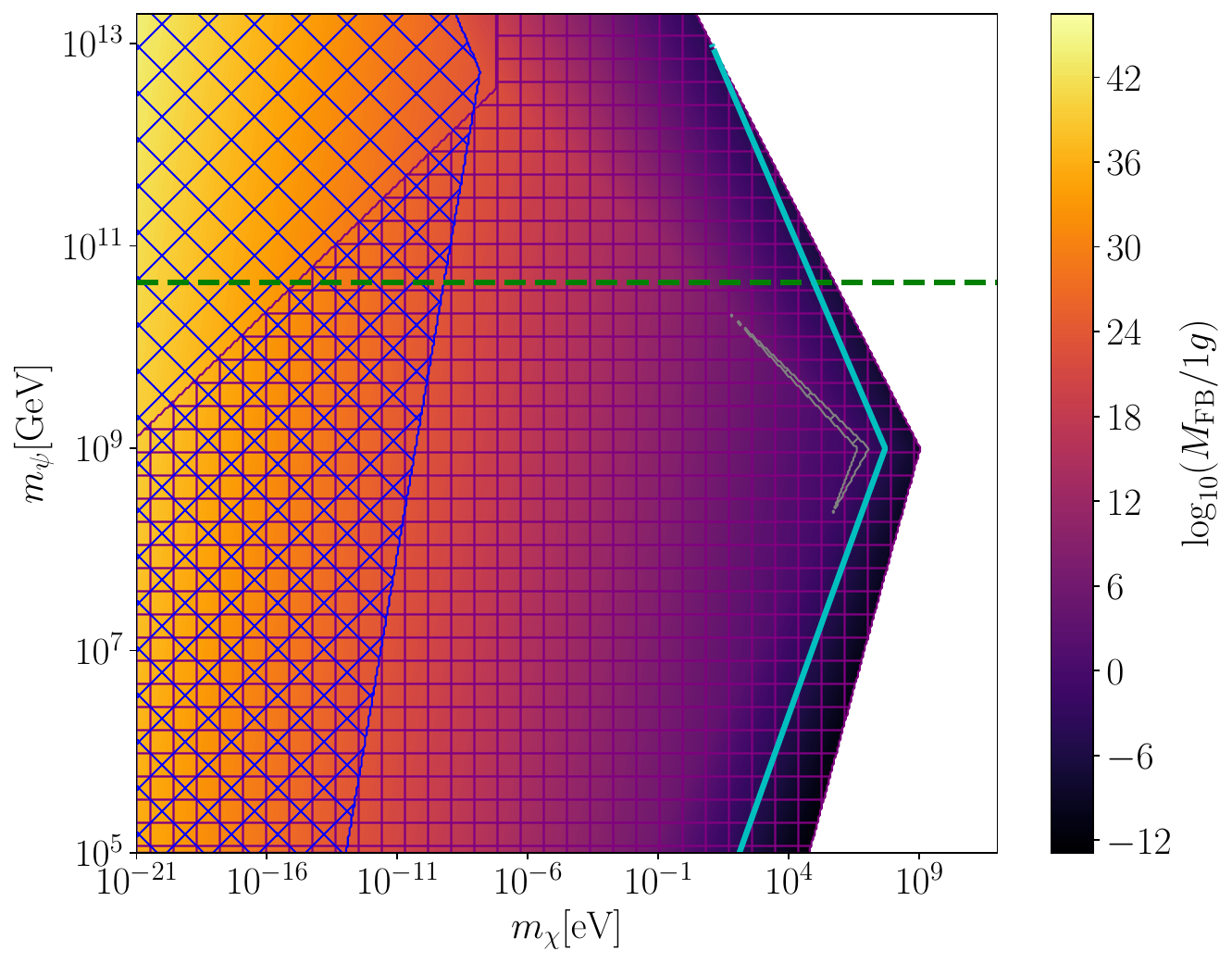}
    \caption{Predicted Fermi ball mass. The green dashed line gives $f_{\rm FB} = 1$. The solid contour in cyan indicates Fermi balls with Planck mass. The blue grids cover the region that is constrained by microlensing and gray region marks where cooling could be faster than mergers. The parameters are chosen to be $\mphi = 10^{16}$ GeV, $y=5\times 10^{-2}$ and $\kappa = 2 \times 10^{-4}$ and $\eta = 10^{-20}$.} 
    \label{FBDM}
\end{figure}

\begin{figure}
    \centering
    \includegraphics[width=0.48\textwidth]{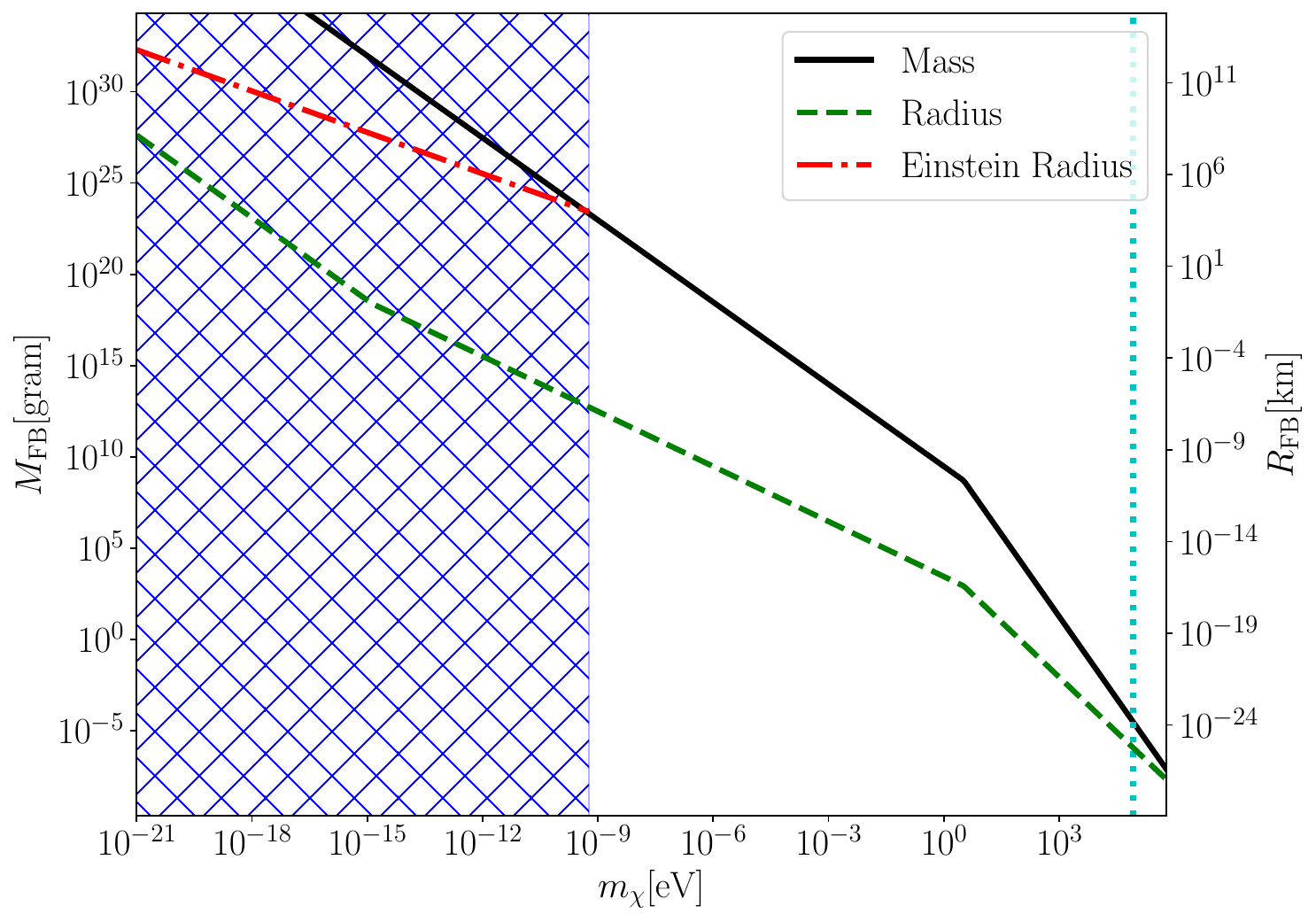}
    \caption{Mass and radius for Fermi balls (or PBHs) with $f_{\rm FB} = 1$ in \figref{FBDM}. The blue meshed region is constrained by microlensing observations. For comparison, the typical Einstein radius in EROS is also shown. The vertical dotted line indicates Planck-massed Fermi balls.} 
    \label{radius}
\end{figure}

Besides PBHs, our model can also accommodate the formation of compact Fermi balls with a wide range of mass that even extend down below the Planck mass. We demonstrate such an example in \figref{FBDM}. Since the Fermi balls only interact with the SM matter through gravity, they also serve as candidate of dark matter. We define $f_{\rm FB}$ as the fraction of DM in Fermi balls in a similar fashion. In \figref{radius}, we show the Fermi ball (or PBH) mass and the radius that correspond to the $f_{\rm FB} = 1$ contour in \figref{FBDM} (green dashed line). In the Fermi balls mass range, their abundance can be constrained from the microlensing observations \cite{Smyth:2019whb, Niikura:2017zjd, Griest:2013esa, Griest:2013aaa, EROS-2:2006ryy, Macho:2000nvd, Oguri:2017ock, Zumalacarregui:2017qqd, Wilkinson:2001vv}. For example, the EROS experiment \cite{EROS-2:2006ryy} monitors the microlensing events of stars in the Magellanic Clouds. The typical Einstein radius of such lensing events for a $10^{-5} M_\odot$ Fermi ball is estimated to be 
\begin{equation}
    R_{E} = \sqrt{4 G M \frac{D_{\rm LS} D_{\rm L}}{D_{\rm LS}+D_{\rm L}}} \sim 10^6 \mathrm{ km},
\end{equation}
where $D_{\rm LS} \sim 50$ kpc is the distance from the source to the lensing object and $D_{\rm L}\sim 5$ kpc is the typical distance from the earth to the lensing object. However, such a Fermi ball in \figref{FBDM} in the region with $f_{\rm FB} \sim 1$ has radius less than $1$ meter. Therefore, the current microlensing constraints also apply to Fermi balls. This rules out the possibility of $f_{\rm FB} = 1$ in the mass range $10^{-10} < M/M_\odot < 10^{9}$ \cite{Carr:2020gox}. The window for Fermi balls to account for all of dark matter remains open below $10^{-10}M_\odot$ (however, see the disputed result regarding femtolensing of $\gamma$-ray bursts \cite{Barnacka:2012bm, Katz:2018zrn}). In particular, it includes the asteroid mass range that is currently unconstrained. It is proposed in Ref. \cite{Das:2021drz} that stellar shocks produced by collision of a star with compact objects like Fermi balls could produce limits in this mass range. On the lower mass end, constraints from PBH evaporation do not apply to Fermi balls as they do not emit particles via Hawking radiation. So the mass window for $f_{\rm FB} = 1$ could extend down to the lowest mass predicted in our model. \textcolor{blue}{}

The Fermi balls behave as noninteracting CDM, even though the dark sector Yukawa attraction is still present. To demonstrate this, we estimate the Fermi ball density at the Galactic center using the generalized Navarro–Frenk–White (NFW) profile
\begin{equation}
    \rho(r) = \frac{2^{3-\alpha} \rho_{0}}{(r / r_s)^{\alpha}(1+ r/r_s)^{3-\alpha}},
\end{equation}
where the scale radius and density are taken to be $r_s = 8$ kpc and $\rho_0 = 1.08 \times 10^{-2} M_\odot \ \rm pc^{-3}$ to match the local value of the DM density. We average this density profile within 1 pc and find the average distance between Fermi balls to be 
\begin{equation}
    \langle d \rangle \sim 
    \begin{cases}
        0.08 (M_{\rm FB} / M_\odot)^{1/3}\ \rm pc \quad  &\alpha = 1\\
        0.004 (M_{\rm FB} / M_\odot)^{1/3}\ \rm  pc \quad &\alpha = 2
    \end{cases}.
\end{equation}
Then the ratio of the average distance to the Yukawa force range can be parametrized by
\begin{equation}
    \langle d \rangle \mchi \sim \zeta \left( \frac{\mpsi}{10^{11} \rm GeV}\right)^{1/3} \left( \frac{\eta}{10^{-20}}\right)^{1/3} \left(\frac{T}{1 \rm GeV}\right),
    \label{distanceratio}
\end{equation}
where $\zeta = 0.7 \times 10^{10}\ (\rm or \ 3.7 \times 10^{8}) \gg 1$ for $\alpha = 1\ (\rm or\ 2)$ is the numerical factor coming from parametrization. Depending on the formation scenario, the temperature $T$ in \equaref{distanceratio} is either $T_g$ or $T_c$. We check that in both cases, $T$ is always greater than 1 GeV. Therefore, the self-interaction between Fermi balls is negligible.

In summary, we described formation of primordial compact objects driven by Yukawa interactions starting from the inflaton decay  and followed by halo formation in the dark sector. There is a wide range of possible masses, and the outcome falls into two distinct formation scenarios. The case of noninstantaneous formation admits a rich merger history and could be better understood with the aid of future numerical simulations that take into account dissipation effects. The end result of the growth of structure is a population of either PBHs or Fermi balls. The  massive PBHs could seed the high redshift AGNs, while, at the lower mass end, the Fermi balls could account for all of the dark matter. Since the constraints for evaporating PBHs do not apply to compact Fermi balls with mass less than $10^{15}$ g, future work is needed to constrain this possible mass range.

\begin{acknowledgments}
We thank Michael S. Turner for helpful discussions.   This work was supported by
the U.S. Department of Energy (DOE) Grant No. DE-SC0009937 and by
the UC Southern California Hub, with funding from the UC National Laboratories division of the University of California Office
of the President.  The work of 
A.K. was also supported  by  the World Premier International Research Center Initiative (WPI),  MEXT,  Japan, and by Japan Society for the Promotion of Science (JSPS) KAKENHI Grant No. JP20H05853. M.M.F was supported by the University of California, Office of the President Dissertation Year Fellowship and donors to the UCLA Department of Physics \& Astronomy.

\end{acknowledgments}

\bibliographystyle{apsrev4-1}
\bibliography{main.bib}

\end{document}